\documentclass[12pt]{article}
\usepackage[dvips]{graphicx}

\renewcommand{\bar}[1]{\overline{#1}}
\newcommand{\half}{{$\frac{1}{2}$}} 

\begin{document}

\begin{flushright}
SLAC-PUB-9135 \\
USM-TH-121 \\

January, 2002
\end{flushright}

\bigskip\bigskip
\begin{center}
{\Large \bf Final-State Interactions  and Single-Spin Asymmetries
in Semi-Inclusive Deep Inelastic
Scattering}\footnote{\baselineskip=13pt Work partially supported
by the Department of Energy, contract DE--AC03--76SF00515, by the
LG Yonam Foundation, and by Fondecyt (Chile) under grant 8000017.}
\end{center}

\vspace{13pt}

\centerline{ \bf Stanley J. Brodsky$^a$, Dae Sung Hwang$^{a b}$,
and Ivan Schmidt$^{c}$}

\vspace{8pt} {\centerline{$^a$Stanford Linear Accelerator
Center,}}

{\centerline{Stanford University, Stanford, California 94309,
USA}}

\centerline{e-mail: sjbth@slac.stanford.edu}

\vspace{8pt} {\centerline{$^{b}$ Department of Physics, Sejong
University, Seoul 143--747, Korea}}

\centerline{e-mail: dshwang@sejong.ac.kr}

\vspace{8pt} {\centerline {$^{c}$Departamento de F\'\i sica,
Universidad T\'ecnica Federico Santa Mar\'\i a,}}

{\centerline {Casilla 110-V, 
Valpara\'\i so, Chile}}

\centerline{e-mail: ischmidt@fis.utfsm.cl }

\vfill

\centerline{Submitted to Physics Letters B.} \vfill
\newpage

\setlength{\baselineskip}{13pt}


\bigskip
\bigskip

\begin{abstract}

Recent measurements from the HERMES and SMC collaborations show a
remarkably large azimuthal single-spin asymmetries $A_{UL}$ and
$A_{UT}$ of the proton in semi-inclusive pion leptoproduction
$\gamma^*(q) p \to \pi X$.  We show that final-state interactions
from gluon exchange between the outgoing quark and the target
spectator system lead to single-spin asymmetries in deep
inelastic lepton-proton scattering at leading twist in
perturbative QCD; {\em i.e.}, the rescattering corrections are not
power-law suppressed at large photon virtuality $Q^2$ at fixed
$x_{bj}$.  The existence of such single-spin asymmetries requires
a phase difference between two amplitudes coupling the proton
target with $J^z_p = \pm {1\over 2}$ to the same final-state, the
same amplitudes which are necessary to produce a nonzero proton
anomalous magnetic moment.  We show that the exchange of gauge
particles between the outgoing quark and the proton spectators
produces a Coulomb-like complex phase which depends on the angular
momentum $L^z$ of the proton's constituents and is thus distinct
for different proton spin amplitudes. The single-spin asymmetry
which arises from such final-state interactions does not factorize
into a product of distribution function and fragmentation
function, and it is not related to the transversity distribution
$\delta q(x,Q)$ which correlates transversely polarized quarks
with the spin of the transversely polarized target nucleon.
\end{abstract}


 \bigskip \bigskip

\section{Introduction}

Single-spin asymmetries in hadronic reactions have been among the
most difficult phenomena to understand from basic principles in
QCD.  The problem has become more acute because of the
observations by the HERMES \cite{hermes0001} and SMC \cite{smc99}
collaborations of a strong correlation between the target proton
spin $\vec S_p$ and the plane of the produced pion and virtual
photon in semi-inclusive deep inelastic lepton scattering $\ell
p^\uparrow \to \ell^\prime \pi X$ at photon virtuality as large as
$Q^2= 6$ GeV$^2$.  Large azimuthal single-spin asymmetries have
also been seen in hadronic reactions such as $p p^\uparrow \to \pi
X$ \cite{E70496}, where the target antiproton is polarized normal
to the pion production plane, and in $ p p \to \Lambda^\uparrow X$
\cite{lambda}, where the hyperon is polarized normal to the
production plane.

In the target rest frame, single-spin correlations correspond to
the $T$-odd triple product $i \vec S_p \cdot \vec p_\pi \times
\vec q,$ where the phase $i$ is required by time-reversal
invariance.  The differential cross section thus has an azimuthal
asymmetry proportional to $|\vec p_{\pi}||\vec q| {\rm sin}
\theta_{q \pi} {\rm sin} \phi$ where $\phi$ is the angle between
the plane containing the photon and pion and the plane containing
the photon and proton polarization vector $\vec S_p.$ In a general
frame, the azimuthal asymmetry has the invariant form ${i\over M}
\epsilon_{\mu \nu \sigma \tau} P^\mu S_p^\nu p^\sigma_\pi q^\tau $
where the polarization four-vector of the proton satisfies $S_p^2 =
-1$ and $S_p \cdot P = 0.$

In order to produce a correlation involving a transversely-polarized
proton, there are two necessary conditions:  (1) There must be two
proton spin amplitudes\\ $M[{\gamma^* p(J^z_p)\to F}]$ with $J^z_p = \pm
{1\over 2}$ which couple to the same final-state $|F>$; and (2) The two
amplitudes must have different, complex phases.
The correlation is proportional to
${\rm Im}(M[J^z_p=+{1\over 2}]^*M[J^z_p=-{1\over 2}])$.
The analysis of
single-spin asymmetries thus requires an understanding of QCD at the
amplitude level, well beyond the standard treatment of hard inclusive
reactions based on the factorization of distribution functions and
fragmentation functions.  Since we need the interference of two
amplitudes which have different proton spin $J^z_p= \pm {1\over 2}$ but
couple to the same final-state, the orbital angular momentum of the two
proton wavefunctions must differ by $\Delta L^z = 1.$ The anomalous magnetic
moment for the proton is also proportional to the interference of
amplitudes $M[{\gamma^* p(J^z_p) \to F}]$ with $J^z_p = \pm {1\over 2}$
which couple to the same final-state $|F>$.

Final-state interactions (FSI) in gauge theory can affect deep
inelastic scattering reactions in a profound way, as has
been demonstrated recently \cite{Brodsky:2001ue}.  The rescattering of the
outgoing quark leads to a leading twist contribution to the deep
inelastic cross section from diffractive channels $\gamma^* p \to
q \bar q p^\prime,$ and the interference effects induced by these
diffractive channels cause nuclear shadowing.  Here we shall show
that FSI also provide the required phase needed to produce
single-spin asymmetries in deep inelastic scattering.

The dynamics of the constituents in the target can be described by its
light-front wavefunctions, $\psi_{n/p}(x_i,\vec k_{\perp i},\lambda_i),$
the projections of the hadronic eigenstate on the free color-singlet
Fock state $|n>$ at a given light-cone time $\tau = t+ z/c.$ The
wavefunctions are Lorentz-invariant functions
of the relative coordinates $x_i= k^+_i/P^+ = (k^0_i + k^z_i)/(P^0+P^z)$
and ${\vec k_{\perp i}}$ [with $\sum^n_{i=1} x_i = 1$ and $\sum^n_{i=1} {\vec
k_{\perp i}} = {\vec 0_\perp}],$ and they are
independent of the bound state's
physical momentum $P^+$ and ${\vec P_\perp}$~\cite{Lepage:1980fj}.   The
physical transverse momenta are ${\vec p_{\perp i}} = x_i {\vec
P_\perp} + {\vec k_{\perp i}}.$ The $\lambda_i$ label the light-front
spin $S^z$ projections of the quarks and gluons along the quantization
$z$ direction.  If a target is stable, its light-front wavefunction must
be real.  Thus the only source of a nonzero complex phase in
leptoproduction in the light-front frame are final-state interactions.
The rescattering corrections from final-state exchange of gauge
particles produce Coulomb-like complex phases which, however, depend on
the proton spin.  Thus $M[{\gamma^* p(J^z_p= \pm {1\over 2} ) \to F}]=
|M[{\gamma^* p(J^z_p=\pm {1\over 2} ) \to F}]|\, e^{i \chi_\pm}$.  Each
of the phases is infrared divergent; however the difference $\Delta \chi
= \chi_+ - \chi_-$ is infrared finite and nonzero.  The resulting
single-spin asymmetry is then proportional to ${\rm sin} \Delta \chi.$

\section{A Model Calculation of Single-Spin Asymmetries in Gauge Theory}

\begin{figure}
\centering
\includegraphics{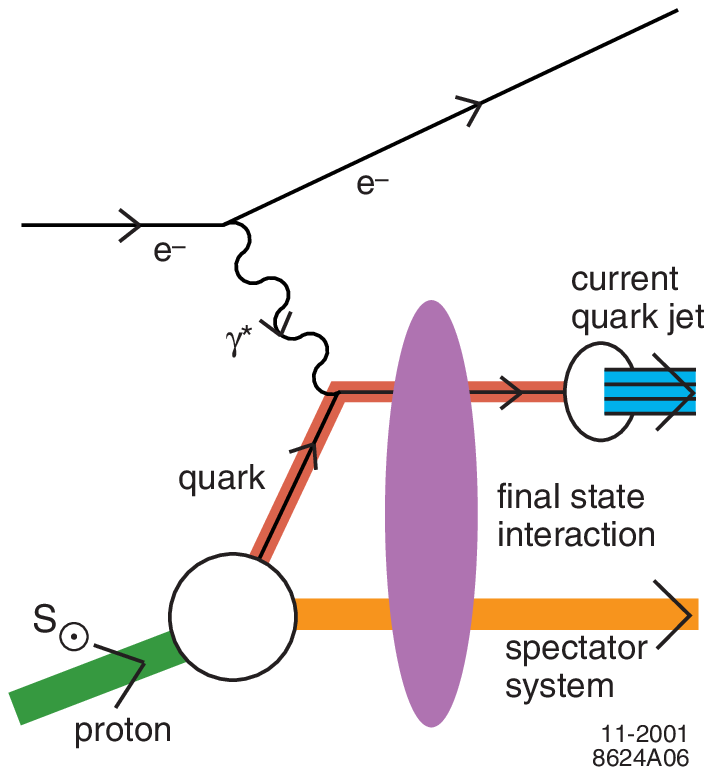}
\caption[*]{The final-state interaction in the semi-inclusive deep
inelastic lepton scattering $\ell p^\uparrow \to \ell^\prime \pi
X$.} \label{fig:1}
\end{figure}

\begin{figure}
\centering \includegraphics{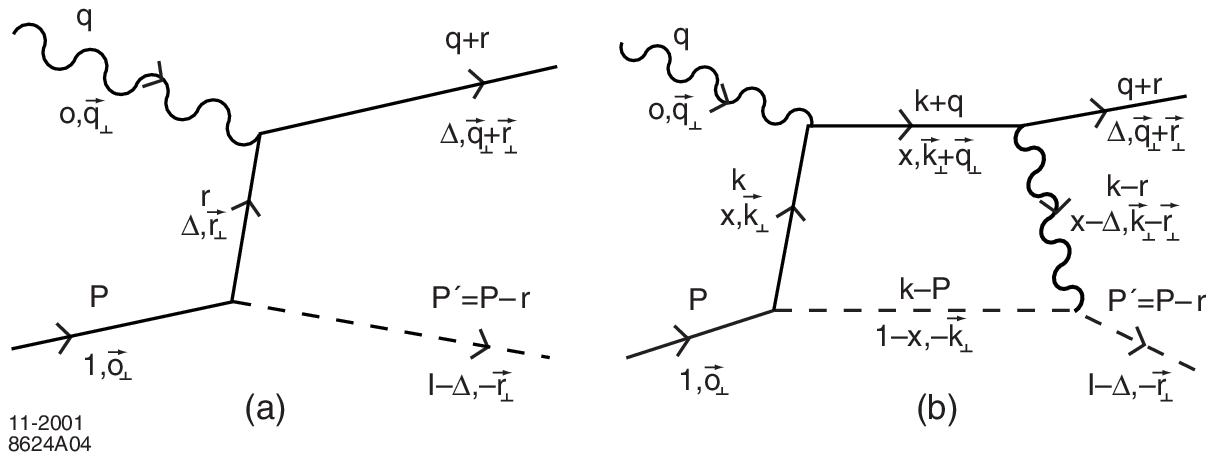}
\caption[*]{The tree (a) and
one-loop (b) graphs for $\gamma^* p \to q (qq)_0$. The
interference of the two amplitudes with $J^z_p = \pm {1/2}$
provides the proton's single-spin asymmetry. } \label{fig:2}
\end{figure}

We shall calculate the single-spin asymmetry in semi-inclusive
electroproduction $\gamma^* p \to H X$ induced by final-state
interactions in a model of a spin-\half ~ proton of mass $M$ with
charged spin-\half and spin-0 constituents of mass $m$ and
$\lambda$, respectively, as in the QCD-motivated quark-diquark
model of a nucleon. The basic electroproduction reaction is then
$\gamma^* p \to q (qq)_0, $ as illustrated in Figs. \ref{fig:1}
and \ref{fig:2}. We shall take the case where the detected
particle $H$ is identical to the quark.  One can take the
asymmetry for a detected hadron by convoluting the jet asymmetry
result with a realistic fragmentation function; {\em e.g.}  $D_{q
\to \pi X}(z,Q^2).$

The amplitude for the $\gamma^* p \to q (qq)_0$ can be computed
from the tree and one-loop graphs illustrated in Fig. \ref{fig:2}.
A spin asymmetry will arise from the final-state interactions of
the outgoing charged lines.
The $J^z = + {1\over 2}$ two-particle
Fock state is given by \cite{BD80,Brodsky:2000ii}
\begin{eqnarray}
&&\left|\Psi^{\uparrow}_{\rm two \ particle}(P^+, \vec P_\perp = \vec
0_\perp)\right>
\label{sn1}\nonumber \\
&=& \int\frac{{\mathrm d}^2 {\vec k}_{\perp} {\mathrm d} x
}{{\sqrt{x(1-x)}}16 \pi^3} \Big[ \ \psi^{\uparrow}_{+\frac{1}{2}}
(x,{\vec k}_{\perp})\, \left| +\frac{1}{2}\, ;\,\, xP^+\, ,\,\,
{\vec k}_{\perp} \right>  \\
&& \qquad +\psi^{\uparrow}_{-\frac{1}{2}} (x,{\vec k}_{\perp})\,
\left| -\frac{1}{2}\, ;\,\, xP^+\, ,\,\, {\vec k}_{\perp} \right>\
\Big]\ , \nonumber
\end{eqnarray}
where
\begin{equation}
\left
\{ \begin{array}{l}
\psi^{\uparrow}_{+\frac{1}{2}} (x,{\vec k}_{\perp})=(M+\frac{m}{x})\,
\varphi \ ,\\
\psi^{\uparrow}_{-\frac{1}{2}} (x,{\vec k}_{\perp})=
-\frac{(+k^1+{\mathrm i} k^2)}{x }\,
\varphi \ .
\end{array}
\right.
\label{sn2}
\end{equation}
The scalar part of the wavefunction $\varphi$ depends on the dynamics.
In the perturbative theory it is simply
\begin{equation}
\varphi=\varphi (x,{\vec k}_{\perp})=\frac{ {e\over \sqrt{1-x}}}{M^2-{{\vec
k}_{\perp}^2+m^2\over x}-{{\vec k}_{\perp}^2+\lambda^2\over 1-x}}\ .
\label{wfdenom}
\end{equation}
In general one normalizes the Fock state to unit probability.

Similarly, the $J^z = - {1\over 2}$ two-particle Fock state has components
\begin{equation}
\left
\{ \begin{array}{l}
\psi^{\downarrow}_{+\frac{1}{2}} (x,{\vec k}_{\perp})=
\frac{(+k^1-{\mathrm i} k^2)}{x }\,
\varphi \ ,\\
\psi^{\downarrow}_{-\frac{1}{2}} (x,{\vec k}_{\perp})=(M+\frac{m}{x})\,
\varphi \ .
\end{array}
\right.
\label{sn2a}
\end{equation}
The spin-flip amplitudes in (\ref{sn2}) and (\ref{sn2a}) have
orbital angular momentum projection $l^z =+1$ and $-1$
respectively.  The numerator structure of the wavefunctions is
characteristic of the orbital angular momentum, and holds for both
perturbative and non-perturbative couplings.

We require the interference between the tree amplitude of Fig.
\ref{fig:2}a and the one loop graph of Fig. \ref{fig:2}b. The
contributing amplitudes for $\gamma^* p \to q (qq)_0$ have the
following structure through one loop order:
\begin{eqnarray}
{\cal A}(\Uparrow \to \uparrow)&=&(M+{m\over \Delta})\ C\
(h+i{e_1e_2\over 8\pi}g_1)
\label{s1}\\
{\cal A}(\Downarrow \to \uparrow)&=&\ ({+r^1-ir^2\over \Delta})\ \ C\
(h+i{e_1e_2\over 8\pi}g_2)
\label{s2}\\
{\cal A}(\Uparrow \to \downarrow)&=&\ ({-r^1-ir^2\over \Delta})\ \ C\
(h+i{e_1e_2\over 8\pi}g_2)
\label{s3}\\
{\cal A}(\Downarrow \to \downarrow)&=&(M+{m\over \Delta})\ C\
(h+i{e_1e_2\over 8\pi}g_1) \ , \label{s4}
\end{eqnarray}
where
\begin{eqnarray}
C&=&-\ g\ e_1\ P^+\ {\sqrt{\Delta}}\ 2\ \Delta\ (1-\Delta)
\label{s5}\\
h&=& {1\over {\vec r}_{\perp}^2+\Delta
(1-\Delta)(-M^2+{m^2\over\Delta} +{\lambda^2\over 1-\Delta})}\ .
\label{s6}
\end{eqnarray}
The quark light-cone fraction $\Delta = {k^+\over P^+}$ is equal
to the Bjorken variable $x_{bj}$ up to corrections of order $1/Q.$
The label $\Uparrow/\Downarrow$ corresponds to $J^z_p = \pm
{1\over 2}.$ The second label $\uparrow/\downarrow$ gives the spin
projection $J^z_q = \pm {1\over 2}$ of the spin-\half~
constituent.  Here $e_1$ and $e_2$ are the electric charges of $q$
and $(qq)_0$, respectively, and $g$ is the coupling constant of
the proton-$q$-$(qq)_0$ vertex.  The first term in (\ref{s1}) to
(\ref{s4}) is the Born contribution of the tree graph. The crucial
result will be the fact that the contributions $g_1$ and $g_2$
from the one-loop diagram Fig. \ref{fig:2}b are different, and
that their difference is infrared finite.  A gauge particle mass
$\lambda_g$ will be used as an infrared regulator in the
calculation of $g_1$ and $g_2.$

The calculation will be done using light-cone time-ordered
perturbation theory, or equivalently, by integrating Feynman loop
diagrams over $dk^-$. The light-cone frame used is $p=
(p^+,p^-,\vec p_\perp)=(P^+, M^2/P^+, \vec 0_\perp)$ and $q =
(q^+, q^-, \vec q_\perp)$ with $q^+ = 0$ and $q^- = 2 q \cdot p
/P^+$, $\vec q_\perp = Q \hat x$ with $Q^2 = -q^2$.  The Bjorken
variable is $\Delta = Q^2/ 2 q \cdot p = Q^2/ 2M \nu.$ Since $q^+
= 0, $ light-cone time-orderings where the virtual photon produces
a $q \bar q$ pair do not appear.

\begin{figure}
\centering \includegraphics{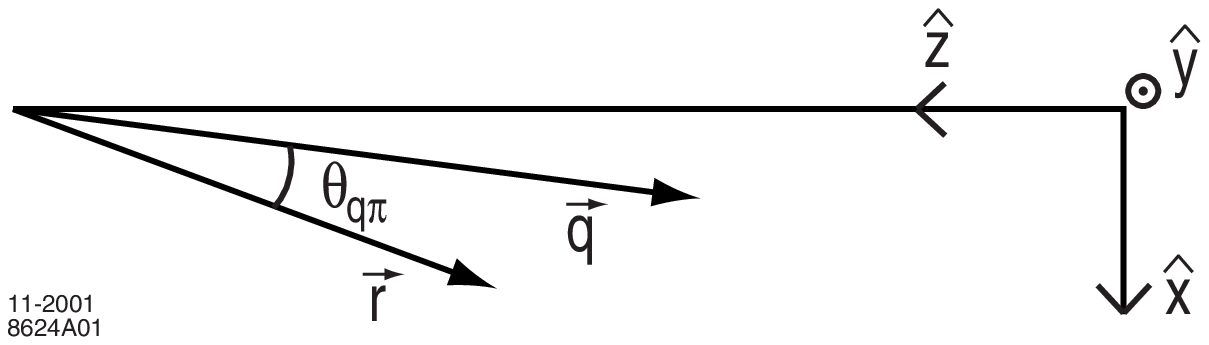} \caption[*]{The light-cone
frame used is $p= (p^+,p^-,\vec p_\perp)=(P^+, M^2/P^+, \vec
0_\perp)$ and $q = (q^+, q^-, \vec q_\perp)$ with $q^+ = 0$. The
virtual photon and produced hadron define the production plane
which we will take as the ${\hat z}-{\hat x}$ plane.}
\label{fig:3}
\end{figure}

The light-cone formalism is invariant under boosts in the $\hat z$
direction: $P^+ \to \gamma P^+$.  It reduces to a laboratory frame
when $P^+ = M$.  If we take $\vec q$ to lie in the $\hat z - \hat x$
plane in this frame, $\vec q = (q^x, q^y, q^z) =
(Q,0,-\nu )$; {\em i.e.}, $\vec q$ is
oriented at an angle $\theta_{lab} = \tan^{-1}{ Q\over
\nu,}$ from the negative $\hat z$ direction.
This is illustrated in  Fig. \ref{fig:3}.
Here $\nu$ is the laboratory energy of the photon.  In the Bjorken
scaling limit with $Q^2$ and $\nu$ large, and $\Delta = x_{bj}$
fixed, the angle $\theta_{lab} \to 0$, so the light-cone
laboratory frame and usual laboratory frame with $\vec q$ taken in
the $-\hat z$ direction are identical.

The covariant expression for the four one-loop amplitudes of
diagram Fig. \ref{fig:2}b is:
\begin{eqnarray}
&&{\cal A}^{\rm one-loop}(I)
\label{fa1s}\\
&=&ig\ e_1^2\ e_2\ \int {d^4k\over (2\pi)^4}
\nonumber\\
&&\times
{{\rm {\cal N}}(I)\over
(k^2-m^2+i\epsilon )\ ((k+q)^2-m^2+i\epsilon )
((k-r)^2-\lambda_g^2+i\epsilon )((k-P)^2-\lambda^2+i\epsilon
)}
\nonumber\\
&=&-ig\ e_1^2\ e_2\ \int {d^2{\vec{k}}_{\perp}\over 2(2\pi)^4}
\int P^+dx\ {{\rm {\cal N}}(I)\over P^{+4}\ x\ x\ (x-\Delta )\ (1-x)}\
\nonumber\\
&&\times
\int dk^- {1\over
\left(k^--{(m^2+{\vec{k}}_{\perp}^2)-i\epsilon\over xP^+}\right)
\left((k^-+q^-)-{(m^2+({\vec{k}}_{\perp}+{\vec{q}}_{\perp})^2)
-i\epsilon\over xP^+}\right)}
\nonumber\\
&&\times
{1\over
\left((k^--r^-)-{(\lambda_g^2+({\vec{k}}_{\perp}-{\vec{r}}_{\perp})^2)
-i\epsilon\over (x-\Delta )P^+}\right)
\left((k^--P^-)+{(\lambda^2+{\vec{k}}_{\perp}^2)-i\epsilon\over
(1-x)P^+}\right)},\
\nonumber
\end{eqnarray}
where we used $k^+=xP^+.$  The numerators ${\rm {\cal N}}(I)$ are given by
\begin{eqnarray}
{\rm {\cal N}}(\Uparrow \to \uparrow)&=&
2P^+{\sqrt{\Delta}}\ x\ (M+{m\over x})\ q^-
\label{s9s}\\
{\rm {\cal N}}(\Downarrow \to \uparrow)&=& 2P^+{\sqrt{\Delta}}\ x\
({+k^1-ik^2\over x})\ q^-
\label{s10s}\\
{\rm {\cal N}}(\Uparrow \to \downarrow)&=& 2P^+{\sqrt{\Delta}}\ x\
({-k^1-ik^2\over x})\ q^-
\label{s11s}\\
{\rm {\cal N}}
(\Downarrow \to \downarrow)&=&
2P^+{\sqrt{\Delta}}\ x\ (M+{m\over x})\ q^-
\ ,
\label{s12s}
\end{eqnarray}
where $q^-={Q^2\over \Delta P^+}={2M\nu\over P^+}\ .$ For the
[current]-[gauge propagator]-[current] factor, in Feynman gauge only the
$-g^{+-}$ term of the gauge propagator $-g^{\mu\nu}$ contributes
in the Bjorken limit, and it provides a factor proportional to
$q^-$ in the numerator which cancels the $q^-$ in the denominator
of the gauge propagator. Therefore the result scales in the
Bjorken limit.

The integration over $k^-$ in (\ref{fa1s}) does not give zero only if
$0 < x < 1$.
We first consider the region $\Delta < x < 1$.
\begin{eqnarray}
&&{\cal A}^{\rm one-loop}(I)
\label{fa1s2}\\
&=&-ig\ e_1^2\ e_2\ \times\ (2\pi i)\ \int {d^2{\vec{k}}_{\perp}\over 2(2\pi)^4}
\int P^+dx\ {{\rm {\cal N}}(I)\over P^{+4}\ x\ x\ (x-\Delta )\ (1-x)}\
\nonumber\\
&&\times
{1\over
\left(P^--{(\lambda^2+{\vec{k}}_{\perp}^2)-i\epsilon\over
(1-x)P^+}
-{(m^2+{\vec{k}}_{\perp}^2)-i\epsilon\over xP^+}\right)
\left(P^--{(\lambda^2+{\vec{k}}_{\perp}^2)-i\epsilon\over
(1-x)P^+}
+q^--{(m^2+({\vec{k}}_{\perp}+{\vec{q}}_{\perp})^2)
-i\epsilon\over xP^+}\right)}
\nonumber\\
&&\times
{1\over
\left(P^--{(\lambda^2+{\vec{k}}_{\perp}^2)-i\epsilon\over
(1-x)P^+}
-r^--{(\lambda_g^2+({\vec{k}}_{\perp}-{\vec{r}}_{\perp})^2)
-i\epsilon\over (x-\Delta )P^+}\right)
},\
\nonumber
\end{eqnarray}
The result is identical to that obtained from light-cone time-ordered
perturbation theory.

The phases $\chi_i$ needed for single-spin asymmetries come from the
imaginary part of (\ref{fa1s2}),
which arises from the potentially real intermediate state allowed before
the rescattering.
The imaginary part of the propagator (light-cone energy denominator) gives
\begin{eqnarray}
-i\pi &\delta& \left( P^--{(\lambda^2+{\vec{k}}_{\perp}^2)\over
(1-x)P^+} +q^--{(m^2+({\vec{k}}_{\perp}+{\vec{q}}_{\perp})^2)
\over xP^+}\right)\nonumber \\
& =& \ -i\pi\ {1\over P^+}\ {\Delta^2\over {\vec q}_{\perp}^2}\
\delta (x\ -\ \Delta \ -\ {\bar{\delta}})\ , \label{sa1}
\end{eqnarray}
where
\begin{equation}
{\bar{\delta}}\ =\ 2\ \Delta\
{{\vec q}_{\perp}\cdot ({\vec k}_{\perp}-{\vec r}_{\perp})\over {\vec
q}_{\perp}^2}
\ .
\label{sa2}
\end{equation}

Since the exchanged momentum $\bar \delta P^+$ is small, the
light-cone energy denominator corresponding to the gauge
propagator is dominated by the $({\vec k}_{\perp}-{\vec
r}_{\perp})^2 + \lambda^2_g \over (x-\Delta)$ term.  This gets
multiplied by $(x-\Delta)$, so only $({\vec k}_{\perp}-{\vec
r}_{\perp})^2 + \lambda^2_g$ appears in the propagator,
independent of whether the photon is absorbed or emitted.  The
contribution from the region $0 < x < \Delta$ thus compliments the
contribution from the region $\Delta < x < 1$.

We can integrate (\ref{fa1s2}) over the transverse momentum using
a Feynman parame\-tri\-zation to obtain the one-loop terms in
(\ref{s1}) to (\ref{s4}).
\begin{eqnarray}
g_1&=&\int_0^1d\alpha\ {1\over \alpha (1-\alpha){\vec r}_{\perp}^2
+\alpha \lambda_g^2 +(1-\alpha)\Delta (1-\Delta)
(-M^2+{m^2\over\Delta}+{\lambda^2\over 1-\Delta})}
\label{s7}\\
g_2&=&\int_0^1d\alpha\ {\alpha\over \alpha (1-\alpha){\vec
r}_{\perp}^2 +\alpha \lambda_g^2 +(1-\alpha)\Delta (1-\Delta)
(-M^2+{m^2\over\Delta}+{\lambda^2\over 1-\Delta})} \ . \label{s8}
\end{eqnarray}
Although not necessary for our analysis, we will assume for
convenience that the final-state interactions generate a phase
when exponentiated, as in the Coulomb phase analysis of QED.  The
rescattering phases $e^{i\chi_i}$ $(i = 1,2)$ with $\chi_i = {\rm
tan}^{-1}({e_1 e_2 \over 8 \pi }{g_i\over h})$ are thus distinct
for the spin-parallel and spin-antiparallel amplitudes. The
difference in phase arises from the orbital angular momentum
$k_\perp$ factor in the spin-flip amplitude, which after
integration gives the extra factor of the Feynman parameter
$\alpha$ in the numerator of $g_2$. Notice that the phases
$\chi_i$ are each infrared divergent for zero gauge boson mass
$\lambda_g \to 0$, as is characteristic of Coulomb phases.
However, the difference $\chi_1 - \chi_2$ which contributes to the
single-spin asymmetry is infrared finite.
We have verified that
the Feynman gauge result is also obtained in the light cone gauge
using the principal value prescription.
The small
numerator coupling of the light-cone gauge particle is compensated
by the small value for the exchanged $l^+ = \bar \delta P^+$
momentum.

The virtual
photon  and produced hadron define
the  production plane which we will take as the ${\hat z}-{\hat x}$ plane.
The azimuthal
single-spin asymmetry transverse to the production plane is given by
\begin{eqnarray}
{\cal P}_y
&=& {e_1e_2\over 8\pi} \
{2\ \Bigl(\ \Delta\, M+m\ \Bigr)\ r^1\over
\Big[\ \Bigl( \ \Delta\, M+m\ \Bigr)^2\ +\
{\vec r}_{\perp}^2\ \Big]}\
\Big[\ {\vec r}_{\perp}^2+\Delta
(1-\Delta)(-M^2+{m^2\over\Delta} +{\lambda^2\over 1-\Delta})\ \Big]
\nonumber\\
&\times&
\
{1\over {\vec r}_{\perp}^2}\
{\rm ln}{{\vec r}_{\perp}^2
+\Delta (1-\Delta)(-M^2+{m^2\over\Delta}+{\lambda^2\over 1-\Delta})\over
\Delta (1-\Delta)(-M^2+{m^2\over\Delta}+{\lambda^2\over 1-\Delta})}\ .
\label{sa2b}
\end{eqnarray}
The linear factor of $r^1 = r^x$ reflects the fact that the single spin
asymmetry
is proportional to ${\vec S}_p \cdot \vec q \times \vec r$ where
${\vec q} \sim - \nu
\hat z$ and
${\vec S}_p = \pm \hat y.$  Here $\Delta = x_{bj}$.

Our analysis can be generalized to the corresponding calculation
in QCD. The final-state interaction from gluon exchange has the
strength ${e_1 e_2\over 4 \pi} \to C_F \alpha_s(\mu^2).$ The scale
of $\alpha_s$ in the ${\overline{MS}}$ scheme can be identified
with the momentum transfer carried by the gluon $\mu^2= e^{-5/3}
({\vec k}_{\perp}-{\vec r}_{\perp})^2$ \cite{Brodsky:1995ds}.
The matrix elements of the proton to its constituents will have
the same numerator structure as the perturbative model since they
are determined by orbital angular momentum constraints.  The strengths of
the proton matrix elements can be normalized by the anomalous
magnetic moment and the total charge. In QCD, $r_\perp$ is the
magnitude of the momentum of the current quark jet relative to the
virtual photon direction.  Notice that for large $r_\perp$, ${\cal P}_y$
decreases as $ \alpha_s(r^2_\perp) x_{bj} M r_\perp \ln r^2_\perp
\over {r}_{\perp}^2 $. The physical proton mass $M$ appears since
it is present in the ratio of the $L_z = 1$ and $L_z = 0$ matrix
elements. This form is expected to be essentially universal.

\section{Model Predictions}

\begin{figure}[htb]
 \centering \includegraphics[height=4in]{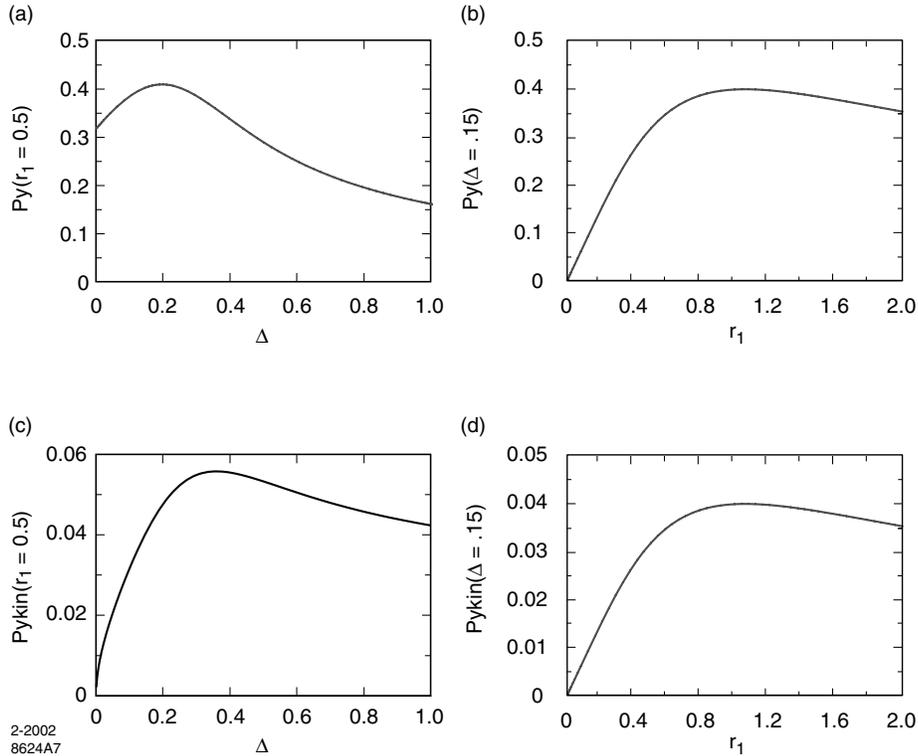} \caption[*]{Model
predictions for the single spin asymmetry of the proton in
electroproduction resulting from gluon exchange in the final state
as a function of $\Delta = x_{bj}$ and quark transverse momentum
$r_\perp$.  The parameters are given in the text.} \label{fig:4}
\end{figure}

 We show the predictions of our model in Fig.~\ref{fig:4} for the
 asymmetry ${\cal P}_y = A^{\sin \phi}_{UT}$
 of the  $i{\vec S}_p \cdot \vec q \times \vec p_q$ correlation
 based on  Eq. ({\ref{sa2b}}).  As representative parameters we take $\alpha_s
 = 0.3$,
 $M =  0.94$ GeV for the proton mass,   $m=0.3$
 GeV for the fermion constituent and $\lambda = 0.8$ GeV for the
 spin-0 spectator.  The single-spin asymmetry ${\cal P}_y$ is shown as a
 function of $\Delta$ at $r_\perp = 0.5$ GeV in Fig. 4a and as a function of
 $r_\perp$ at $\Delta = 0.15$ in Fig. 4b. The Hermes asymmetry
 $A_{UL}^{\sin \phi}$ contains a
 kinematic factor $K = {Q\over \nu}\sqrt{1-y} =
{\sqrt{2Mx\over E}}{\sqrt{1-y\over y}}$ because the proton is
 polarized along the direction of the incident electron. The
 resulting predictions for $K {\cal P}_y$ are shown in Figs. 4c and 4d.
 Note that $\vec r = \vec p_q  - \vec q$ is the momentum of the current
 quark jet relative to the photon direction.  The asymmetry as a
 function of the pion
 momentum $\vec p_\pi$ requires a convolution with the quark fragmentation function.

\section{Summary}

We have calculated the single-spin asymmetry in semi-inclusive
electroproduction induced by final-state
interactions. We have shown that the final-state interactions
from gluon exchange between the outgoing quark and the target
spectator system leads to single-spin asymmetries in deep
inelastic lepton-proton scattering at leading twist in
perturbative QCD; {\em i.e.}, the rescattering corrections are not
power-law suppressed at large photon virtuality $Q^2$ at fixed
$x_{bj}$. The azimuthal single-spin asymmetry ${\cal
P}_y$ transverse to the photon-to-pion production plane decreases
as $ \alpha_s(r^2_\perp) x_{bj} M r_\perp [\ln r^2_\perp]/
{r}_{\perp}^2$ for large $r_\perp,$ where $r_\perp$ is the
magnitude of the momentum of the current quark jet relative to the
virtual photon direction. The  fall-off in $r^2_\perp$
instead of  $Q^2$  compensates for the dimension of the $\bar q$-$q$
-gluon correlation. The mass $M$ of the physical proton mass appears
here since it determines the ratio of the $L_z = 1$ and $L_z = 0$
matrix elements. We  have estimated the scale of $\alpha_s$ as
${\cal O}( r_\perp^2).$
The nominal size of the spin asymmetry is thus $C_F
\alpha_s(r^2_\perp) a_p$ where $a_p$ is the proton anomalous
magnetic moment.

It is usually assumed that the cross section for semi-inclusive
deep inelastic scattering at large $Q^2$ factorizes as the product
of quark distributions times quark fragmentation functions
\cite{collins93,boermulders}. Our analysis shows that the
single-spin asymmetry which arises from final-state interactions
does not factorize in this way since the result depends on the $<p
|\bar \psi_q A \psi |p>$ proton correlator, not the usual quark
distribution derived from $<p |\bar \psi_q (\xi) \psi_q(0)|p>$
evaluated at equal light-cone time $\xi^+ = 0$. In particular, the
spin asymmetry is not related to the transversity distribution
$\delta q(x,Q)$ which correlates transversely polarized quarks
with the spin of the transversely polarized target nucleon.

Our results are directly applicable to the
azimuthal correlation of the proton spin with the virtual photon
to current quark jet plane, which can be deduced from jet measures
such as the thrust distribution. The $\sin \phi$ correlation of
the proton spin with the photon-to-pion production plane as
measured in the HERMES and SMC experiments can then be obtained
using the usual fragmentation function.  Detailed comparisons with
experiment will be presented elsewhere.
Our approach can also be applied to single-spin asymmetries in
more general hadronic hard inclusive reactions such as $e^+ e^-
\to \Lambda^{\uparrow} X$ and $p p \to \Lambda^{\uparrow} X.$

\noindent
{\large \bf Acknowledgments}

\noindent
We thank John Collins, Paul Hoyer, and Stephane Peigne
for helpful comments.

\end{document}